# Large Scale Structure Tests
# of Warm Dark Matter

Stéphane Colombi[1], Scott Dodelson[1] and Lawrence M. Widrow[2]

[1] *NASA/Fermilab Astrophysics Center*
*Fermi National Accelerator Laboratory, Batavia, IL 60510-0500*

[2] *Department of Physics*
*Queen's University, Kingston, Canada, K7L 3N6*

## ABSTRACT

The nature of the dark matter critically affects the large scale structure of the Universe. Under the assumptions that the Universe is spatially flat with zero cosmological constant and that primordial perturbations were adiabatic with a Harrison-Zel'dovich spectrum, neither hot (HDM) nor cold dark matter (CDM) appears consistent with the observed large scale structure. Warm dark matter (WDM) is an intriguing alternative from the point of view of both cosmology and particle physics.

We consider a one-parameter family of WDM models. The linear power spectra for these models is calculated and compared with the corresponding spectra for CDM, HDM and mixed dark matter (MDM) as well as the power spectrum derived from observations. Our linear analyses suggest that a model universe dominated by a particle whose mass to temperature ratio $m_x/T_x$ is increased by a factor of two as compared with the standard HDM neutrino gives a reasonable fit to the data on large ($> 8h^{-1}$ Mpc) scales.

$N$-body simulations for this particular WDM model show features of both HDM and CDM. As in HDM, the first objects to collapse are large pancake-like structures. The final matter distribution is rather smooth and structures as small as galaxy halos are excluded. However, there appear to be virialized rich clusters evident in the CDM but not the HDM simulations. Unfortunately, a simple comparison of the matter distribution and its statistical properties with observations indicates that WDM, like CDM, has too much power at small scales. This is particularly evident in the small-scale pairwize velocity dispersion. The cluster multiplicity function has the wrong shape with too many rich clusters being produced, though this conclusion is based on the simple assumption that light traces mass in groups of galaxies.

# 1 Introduction

While there is ample evidence for dark matter in our Universe, its nature remains a mystery. Is this matter in the form of baryons, massive neutrinos, or something new and exotic? The answer to this question critically affects our understanding of the early Universe and in particular the formation of structures such as galaxies, clusters, and voids.

For the purposes of structure formation, it is the distribution of the dark matter particles in velocity space that is most important. For example, in a cold dark matter (CDM)-dominated universe, the velocity dispersion of the dark matter at the time of matter-radiation equality ($t_{\rm eq}$) is negligible and structure formation begins with the collapse of relatively small objects. Larger mass objects form by aggregation leading to a bottom-up scenario. Hot dark matter (HDM) has large velocity dispersion at $t_{\rm eq}$ and leads to a scenario in which large pancake-shaped objects form first and then fragment into smaller objects (top-down scenario).

HDM and CDM represent extremely simple models in that once one specifies the density of the dark matter, the velocity-space distribution function $f(v)$ is fixed. Of course, to fully specify a cosmological model, one must include the total density ($\rho = \Omega \rho_{\rm crit} = 1.05\,\Omega h^2 \times 10^4\,{\rm eVcm^{-3}}$), the baryon density ($\rho_B = \Omega_B \rho_{\rm crit}$), the Hubble constant today ($H_0 = 100\,h\,{\rm km\,s^{-1}\,Mpc^{-1}}$), the cosmological constant ($\Lambda$), and the initial power spectrum of density perturbations. (Here and throughout, we set $\hbar = c = k_B = 1$.) The "standard" HDM and CDM models have $\Omega = 1$, $\Lambda = 0$, $0.5 < h < 1.0$, $0.01 < \Omega_B < 0.1$ and adiabatic primordial perturbations with $P(k) \propto k$. It now appears that neither of these standard models are consistent with the observations. CDM for example, has too little power on large ($\gtrsim 30\,h^{-1}\,{\rm Mpc}$) scales relative to small ($\lesssim 10\,h^{-1}\,{\rm Mpc}$) scales. HDM, on the other hand, has trouble forming galactic scale structures early enough to be in agreement with observations of high redshift quasars.

One set of alternatives involves nonstandard HDM or CDM scenarios. For example, Albrecht & Stebbins (1992) have shown that wakes of cosmic strings can seed small-scale structures in an HDM-dominated universe thereby avoiding the problems of early galaxy formation. Other possibilities include nonzero $\Lambda$ (Peebles 1984; Turner, Steigman, & Krauss 1984; Efstathiou et al. 1990; Turner 1991), primordial perturbations with a tilted spectrum (i.e., $P(k) \propto k^n$; $n \neq 1$) (Adams et al. 1993), decaying particles (Bond & Efstathiou 1991; Dodelson, Gyuk, & Turner 1994) and mixed hot and cold dark matter (MDM) (Shafi & Stecker 1984; Davis, Summers, & Schlegel 1992; Taylor and Rowan-Robinson 1992; van Dalen & Schaefer 1992; Klypin et al. 1993).

Here, we consider warm dark matter (WDM) cosmologies with $\Omega = 1$, $\Lambda = 0$, $0.5 < h < 1.0$, $\Omega_B = 0$, and primordial perturbations $P(k) \propto k$. (We discuss our choice of $\Omega_B$ below.) By warm dark matter, we mean any particle whose velocity dispersion during the time of structure formation is non-negligible but less than the velocity dispersion for standard HDM.

To keep things simple, we consider a one-parameter family of distribution functions for the dark matter candidate which interpolate between the distribution functions for HDM and CDM. To be precise, we take the distribution function for the dark matter or "$x$" particles to be

$$f_x(v) = \frac{\beta}{e^{p/\alpha T_\gamma} + 1} \qquad (1)$$

where $T_\gamma$ is the photon temperature, $v = p/\left(p^2 + m_x^2\right)^{1/2}$ and $m_x$ is the particle's mass. The distribution function is specified by three parameters $\alpha$, $\beta$, and $m_x$. However, for the purposes of understanding structure formation, only two combinations of these are relevant, one related to $\Omega_x h^2$ and the other related to the shape of the distribution function. In standard HDM, $\alpha = (4/11)^{1/3}$, $\beta = 1$, and the remaining parameter – the mass – is chosen to set $\Omega$. This leaves no freedom for the shape. In CDM scenarios, the velocity dispersion is negligible and therefore the actual form of the distribution function is irrelevant. For our purposes, it is useful to think of CDM particles as having a distribution function given by Eq. (1) in the limiting case $\alpha = $ constant, $\beta \to 0$, and $m \to \infty$. (Equivalently, we can keep $\beta$ fixed and let $\alpha \to 0$ and $m \to \infty$.) For the family of WDM models considered here, $\alpha$ and/or $\beta$ vary from their canonical HDM value. The models therefore have one additional degree of freedom as compared with standard HDM or CDM and by varying this parameter, one interpolates between CDM and HDM. The remaining parameter describes a family of models that are equivalent from the point of view of large scale structure though distinct in terms of how the dark matter particles were produced. These points will be discussed in detail in Section 2.

This work is, at least in spirit, similar to that done for MDM. MDM models contain an admixture of hot and cold particles and can also be described as a one parameter family which smoothly interpolates between HDM and CDM. But as we will see, there are both qualitative and quantitative differences between MDM and WDM cosmologies.

WDM, along with CDM, was introduced in the early 80's (Pagels & Primack 1982; Peebles 1982; Bond, Szalay & Turner 1982, Olive & Turner 1982) when it became clear that HDM had serious flaws. CDM has of course received far more attention and for good reason. First, WDM, with an additional free parameter, is less predictive. Second, the early candidates for WDM were not particularly compelling in that they required a new particle in the $100\,\text{eV} - 1\,\text{keV}$ range, well within the reach of particle accelerators. However, both of these reasons have become obsolete. First, as mentioned previously, the standard CDM model does not seem to fit the data and so models with more freedom are now in vogue. Second, a better understanding of the early Universe has led to a number of WDM candidates such as right-handed or sterile neutrinos suggesting that, at least from the point of view of particle physics, WDM is as palatable as CDM.

The rest of the paper focuses on understanding large scale structure in a WDM-dominated universe and comparing the results with observations. We begin with linear perturbation theory. In Section 3, we outline our calculation of the linear transfer function and discuss, in section 4, various tests using the derived power spectra. The strategy is to use linear tests to survey the family of WDM models and determine which is most promising. We also use this opportunity to compare these models with the other possibilities such as MDM. We conclude that large scale structure in a universe dominated by a particle whose mass to temperature ratio $m_x/T_x$ is roughly twice that of the standard HDM is in reasonably good agreement with the data. Linear theory also suggests that there are problems with early galaxy formation though here, we are in the non-linear regime and so should use caution before reaching any conclusions. Proceeding to the next level of approximation, we carry out detailed N-body simulations of a model WDM-dominated universe and compare with similar simulations for CDM and HDM. The results are discussed in Section 5. In particular, we visually analyse large scale structures, we study the (non-linear) power spectrum, the two-point correlation function, pairwise velocities, and the group multiplicity function. A summary and some conclusions are given in Section 6.

## 2 Models of Warm Dark Matter

In this section we motivate two prototype WDM candidates and show that they are equally well described by Eq. (1). First however, we review the standard HDM neutrino.

## 2.1 Hot Dark Matter

The three neutrinos in the Standard Model interact with ordinary matter via the weak interactions. As such they decouple from the primeval electromagnetic plasma at temperatures of order a few MeV and therefore, unlike the photons, are not heated when $e^{\pm}$ annihilate. To calculate the temperature and number density of neutrinos (Weinberg 1972; Kolb & Turner 1990) we first note the Universe expands adiabatically so that the entropy density

$$s = \frac{2\pi^2}{45} g_*(T) T^3 \qquad (2)$$

scales as $a^{-3}$. Here, $a$ is the Robertson-Walker scale factor, $T$ is the common temperature of all particles thermally coupled to the photons, and $g_*(T)$ is the effective number of degrees of freedom of massless particles. After the neutrinos decouple, their temperature, $T_\nu$, scales as $a^{-1}$ and therefore $s/T_\nu^3 = (2\pi^2/45) g_*(T_\gamma/T_\nu)^3$ remains constant. Prior to $e^{\pm}$ annihilation, $g_* = 11/2$ (counting photons, electrons, and positrons) whereas after $e^{\pm}$ annihilation $g_* = 2$. Therefore, $T_\nu/T_\gamma = (4/11)^{1/3}$ and the velocity-space distribution function is

$$f(p) = \frac{1}{e^{p/T_\nu} + 1} \ . \qquad (3)$$

That is, the distribution function is described by Eq.(1) with $\alpha = (4/11)^{1/3}$ and $\beta = 1$. By integrating Eq.(3) over all momenta, one recovers the well-known result (Gerstein & Zel'dovich 1966; Cowsik & McClelland 1972; Marx & Szalay 1972):

$$\Omega_\nu h^2 = \frac{m_\nu}{93 \text{ eV}} \ . \qquad (4)$$

## 2.2 Early-Decoupled Particles

The above results can be generalized to any particle which decouples when it is still relativistic. For particles decoupling earlier than the standard model neutrinos

$$\frac{T_x}{T_\gamma} = \left(\frac{4}{11}\right)^{1/3} \left(\frac{10.75}{g_*(T_D)}\right)^{1/3} \qquad (5)$$

where $T_D$ is the temperature of the Universe when the "$x$" particles decouple. $g_*$ here includes contributions for the three standard model neutrinos (in contrast with the $g_*$ of (2.1)) and is equal to 10.75 for 100 MeV $\lesssim T_D \lesssim$ 1 MeV and 106.75 for $T_D \gtrsim$ 300 GeV (Kolb and Turner 1990). The distribution function for a particle which decouples when $g_* \gtrsim 11$ will have both a lower temperature and lower number density relative to the standard HDM neutrino; that is, $\alpha < (4/11)^{1/3}; \beta = 1$. This in turn implies that for fixed $\Omega_x h^2$, the particle will have a higher mass and therefore reduced velocity dispersion relative to standard HDM. WDM of this type was discussed by Peebles (1982), Bond & Szalay (1983), and Bond, Szalay, & Turner (1982). At that time, the favored WDM candidate was the gravitino, the supersymmetric partner to the graviton.

## 2.3 Right-Handed Neutrinos

Another group of WDM candidates are the right-handed neutrinos. In the standard model, all fermions except the neutrinos have both left and right chiral projections. This is at least in part why neutrinos in the standard model are massless. Right-handed neutrinos (one species for each ordinary neutrino type) are arguably the most natural additions to the standard model. Once right-handed neutrinos are added there is the possibility for Dirac-type neutrino mass terms similar to the terms which give rise to masses for the charged leptons and quarks. In addition, because neutrinos are electrically neutral, there is also the possibility for Majorana mass terms, and therefore oscillations between right and left-handed neutrinos. Oscillations of this type have been invoked in an MSW- (Mikheyev & Smirnov 1986; Wolfenstein 1978) type solution to the solar neutrino problem (Barger *et al.* 1991; Butler and Malaney 1992).

Right-handed neutrinos do not interact via the strong, electromagnetic, or weak interactions and so it is natural to think of them as having been in equilibrium early on and decoupling at relatively high temperatures. If for example, they decouple before the electroweak phase transition ($g_* \sim 100$) then the number density, which scales as $T^3$, will be a factor of ten smaller than that of standard neutrinos. To close the Universe one would therefore need a right-handed neutrino with a mass $m_x \simeq 900h^2$ eV, thereby making it a perfect warm dark matter candidate.

There are two possible problems with the above arguments, one from astrophysics and the other from particle physics. First, as we will see in later sections, a keV mass particle leads to phenomenology very similar to that of CDM, especially on the largest scales. (With this in mind, Malaney, Starkman, and Widrow (1995) have considered MDM models with a right-handed 1 keV neutrino as the cold component and an ordinary neutrino as the hot component. See also Valdarnini & Bonometto 1985.) Second, it was observed by Langacker (1989) that there is no reason to expect right-handed neutrinos to be in equilibrium at early times. In fact, an accurate calculation of the rate for producing right-handed neutrinos indicates that the dominant production mechanism is the oscillation mentioned above. The oscillation rate peaks at temperatures $\sim 100$ MeV suggesting that the number of right-handed neutrinos prior to the electroweak phase transition was negligible. (This calculation has evolved over the years starting with the work of Dolgov (1981). Manohar (1987) presented an interesting model which explained very nicely the quantum mechanics involved. As far as we know, Langacker's work was the first to derive realistic cosmological limits on the various neutrino parameters. Subsequent refinements were introduced by Barbieri & Dolgov (1990, 1991); Enqvist, Kainulainen, & Maalampi (1990a,b); Enqvist, Kainulainen, & Thomson (1992); Cline, (1992).)

Dodelson and Widrow (1994) considered the possibility that a nonequilibrium distribution of neutrinos could be produced by oscillations. In particular, they showed that as long as $g_*$ is constant during the epoch when the neutrinos are produced, their distribution function is given by Eq. (1) with $\alpha = (4/11)^{1/3}$ and $\beta < 1$ where the value of $\beta$ depends on the parameters of the neutrino mass matrix. For fixed $\Omega_x$, decreasing $\beta$ corresponds to increasing the mass.

## 2.4 Distribution Functions

The generic WDM candidate therefore has a distribution function given by Eq. (1) with three parameters, $\alpha$, $\beta$ and $m$. Fixing the density of the particles implies one constraint:

$$\Omega_x h^2 = \beta \ \left(\frac{\alpha^3}{(4/11)}\right) \left(\frac{m_x}{93 \text{ eV}}\right). \qquad (6)$$

This leaves two free parameters, which we can choose to be $m_x/\alpha$ and $\alpha$. The former is proportional to $m_x/T_x$ and governs the shape of the power spectrum. The remaining parameter $\alpha$ generates a family of models that are equivalent from the point of view of structure formation though distinct if one is interested in how the particles are produced. In particular, for fixed $\Omega_x h^2$ and $m_x/\alpha$, one value of $\alpha$ corresponds to early-decoupled matter and another corresponds to oscillation-produced sterile neutrinos though both lead to exactly the same predictions for large scale structure. Quantitatively, we have

$$\text{Power Spectrum}_{\text{early-decoupled matter}}(m_1) = \text{Power Spectrum}_{\text{sterile neutrinos}}(m_2) \quad (7)$$

where

$$m_2 = 163 \left(\frac{m_1}{100 \text{eV}}\right)^{4/3} \left(\frac{0.5}{h}\right)^{2/3} \text{eV}. \quad (8)$$

To close this section we mention two final points about WDM candidates. Recently Babu, Rothstein, & Seckel (1993) have proposed Majorons as another WDM candidate. Presumably this candidate would have a distribution function like early-decoupled matter. Finally, the distribution function we have taken for sterile neutrinos assumes that $g_*$ is constant during the time when the neutrinos are produced. While this is not always a good assumption, a preliminary analysis of models with a time-dependent $g_*$ does not yield transfer functions terribly different from the ones considered here.

## 3 The Power Spectrum

The growth of perturbations in the early Universe is governed by the Einstein equations coupled to a Boltzmann equation for each type of matter present. Our model Universe consists of three components: ordinary matter (photons, baryons, and electrons), massless, standard model neutrinos, and massive right-handed neutrinos. At early times, the fluctuations in the matter fields are small and one can use linear perturbation theory (Peebles 1982; Bond and Szalay 1983) where the zeroth order solution describes an Einstein-de Sitter universe. In linear theory, the line element can be written

$$ds^2 = dt^2 - a^2(t) \left(\delta_{\alpha\beta} - h_{\alpha\beta}(\mathbf{x}, t)\right) dx^\alpha dx^\beta . \quad (9)$$

The baryon/photon/electron mix is treated as a tightly coupled ideal fluid characterised by a density field $\rho_\gamma$ and a velocity field $v_\gamma$. To first order, the density field can be written:

$$\rho_\gamma(\mathbf{x}, t) = \rho_{\gamma,0}(t) \left(1 + \delta_\gamma(\mathbf{x}, t)\right). \quad (10)$$

This one-fluid approximation greatly simplifies the numerics. While it is valid prior to recombination, a more careful treatment is required if one is interested in small angular scale microwave background distortions and/or if baryons play an important role in the post-recombination evolution of the density perturbations. We leave microwave background calculations for future work. So we are implicitly assuming that $\Omega_B \simeq 0$. This may be in conflict with big bang nucleosynthesis (Copi, Schramm, & Turner 1995 and references therein) and in this respect our models are not as realistic as they could be. However, the differences between the power spectra of $\Omega_B = 0$ models and those with a more realistic $\Omega_B = 0.02 - 0.1$ should be no more than 10%.

We assume that there are three massless neutrino species and one massive neutrino species. To first order, their distribution functions can be written:

$$f_i(\mathbf{p}, \mathbf{x}, t) = f_{i,0}(p, t) - p \frac{\partial f_{i,0}}{\partial p} \Delta_i(\mathbf{p}, \mathbf{x}, t) \quad (11)$$

where $i = \nu, x$ denotes the type of neutrino, $p \equiv |\mathbf{p}|$, and $f_{i,0}$ are the zeroth order distribution functions, given by Eq.(1) with appropriate choices for $\alpha$ and $\beta$.

In the synchronous gauge, the metric perturbations are encapsulated in the two functions $h_{33}$ and $h \equiv \text{Tr}(h_{\alpha\beta})$. $h$, $h_{33}$, $\delta_\gamma$, $v_\gamma$, $\Delta_\nu$, and $\Delta_x$ form a complete set of variables. We expand each in terms of its Fourier components (e.g., $\tilde{\delta}_\gamma(\mathbf{k}, t) = \int d^3 x e^{i\mathbf{k}\cdot\mathbf{x}} \delta_\gamma(\mathbf{x}, t)$). The equations (with the tilde omitted for convenience) are (Peebles 1982, Bond & Szalay 1983):

$$\dot{\Delta}_x + ik\mu \frac{p}{E(p)} \Delta_x = \dot{h}(1 - \mu^2) + \dot{h}_{33}(3\mu^2 - 1) \tag{12}$$

$$\dot{\Delta}_\nu + ik\mu \Delta_\nu = \dot{h}(1 - \mu^2) + \dot{h}_{33}(3\mu^2 - 1) \tag{13}$$

$$\dot{\delta}_\gamma + \frac{4}{3} ikv = \frac{2}{3}\dot{h} \tag{14}$$

$$\dot{v} + \frac{ik\delta_\gamma}{4} = 0 \tag{15}$$

$$\ddot{h} + \frac{\dot{a}}{a}\dot{h} = 16\pi G a^2 \left( \rho_{\gamma,0} \delta_\gamma + \frac{1}{2} \sum_{i=\nu,x} g_i \int \frac{d^3 p}{(2\pi)^3} \left[ E(p) + \frac{p^2}{E(p)} \right] \left( -p \frac{\partial f_{0,i}}{\partial p} \right) \Delta_i \right) \tag{16}$$

$$\dot{h}_{33} - \dot{h} = \frac{16\pi G a^2}{ik} \left( \frac{4}{3} \rho_{\gamma,0} v_\gamma + \sum_{i=\nu,x} g_i \int \frac{d^3 p}{(2\pi)^3} p\mu \left( -p \frac{\partial f_{0,i}}{\partial p} \right) \Delta_i \right). \tag{17}$$

Some notation: $E(p) \equiv \sqrt{p^2 + m^2}$ where $m$ is zero for the massless neutrinos and $m = m_x$ for massive neutrinos; $G$ is Newton's constant; $g_i$ is the number of degrees of freedom for the $i$'th species (equal to two for all the particles here); $\mu = \hat{k} \cdot \hat{p}$ and dot denotes differentiation with respect to conformal time $\tau = \int dt/a(t)$.

The power spectrum today $|\delta\rho_x/\rho_x|^2$ can be expressed as an integral over $\Delta_x(p)$:

$$P(k) = \left| \frac{1}{2\rho_x} \int \frac{d^3 p}{(2\pi)^3} E(p) p \frac{\partial f_0}{\partial p} \Delta_x(\tau_0, k, p) \right|^2, \tag{18}$$

where $\tau_0$ is the conformal time today. Actually, on very large scales ($k \rightarrow 0$) the power spectrum is independent of the type of dark matter present and depends only on the initial perturbations. It is therefore convenient to define the transfer function

$$T(k) = \frac{\frac{\delta\rho_x}{\rho_x}(k)}{\frac{\delta\rho_x}{\rho_x}(k \rightarrow 0)} \tag{19}$$

where by construction, $T(k) \rightarrow 1$ for $k \rightarrow 0$. The power spectrum can then be written

$$P(k) = Bk^n T^2(k) \tag{20}$$

where $n$ is the spectral index for the primordial perturbations, and $B$ is the normalization constant.

The transfer functions for a representative sample of WDM models are shown in top panel of figure 1. Our models all assume $h = 0.5$, $\Omega_B = 0.0$ and $\Omega = 1.0$. For definiteness, we label the models by the mass the neutrinos would have assuming they are produced through oscillations. We define $m_0 = 23$ eV to be the mass of a standard HDM particle in such a universe. The model labelled $2m_0$ therefore refers to a universe dominated by a 46 eV particle whose distribution function is given by Eq. (1) with $\alpha = (4/11)^{1/3}$ and $\beta = 0.5$.

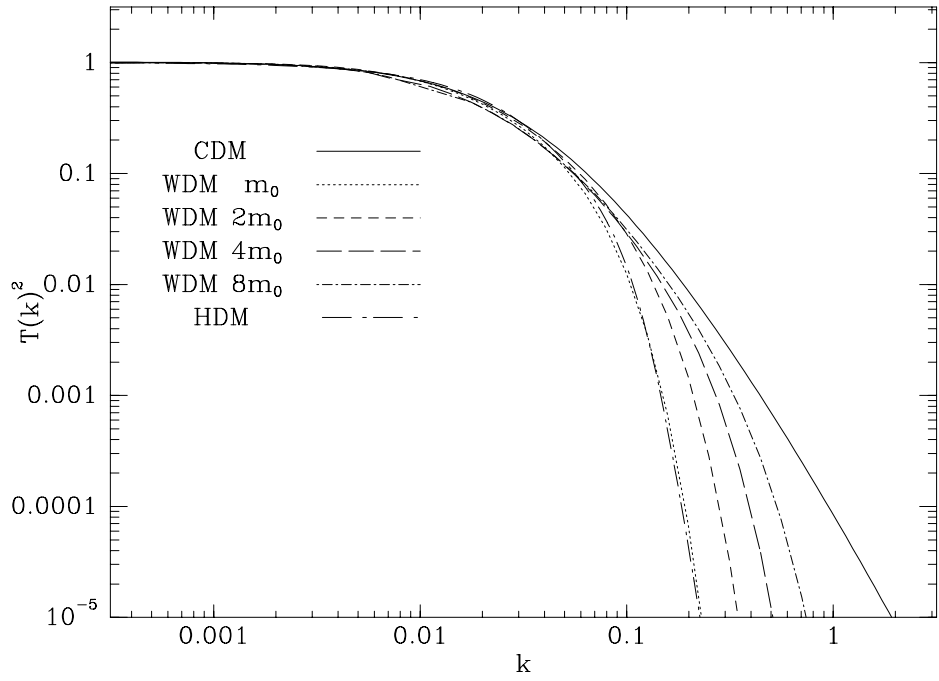

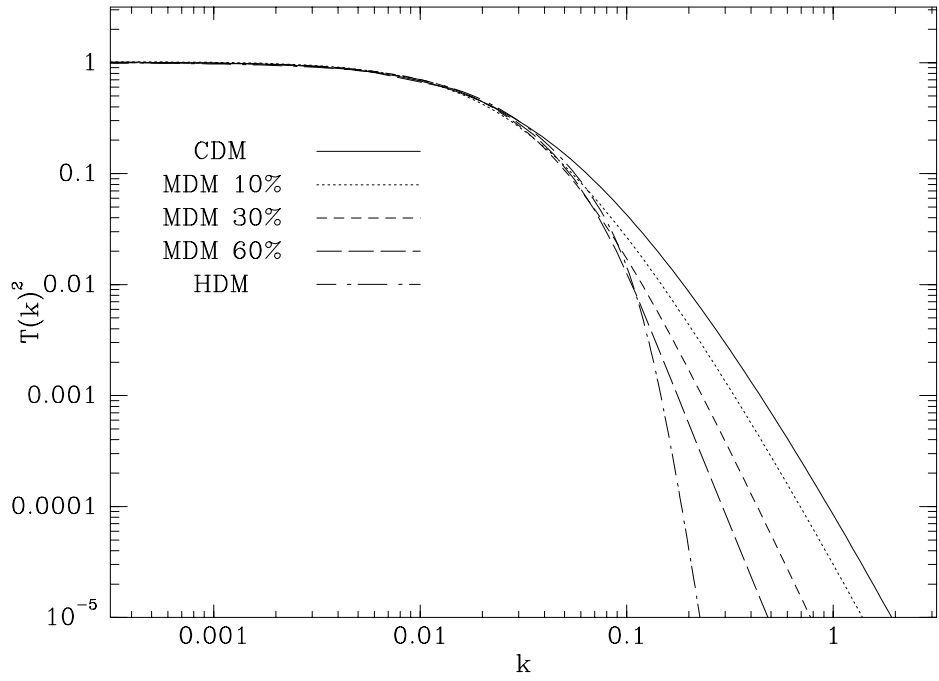

Figure 1: Transfer functions for WDM (top) and MDM (bottom) models. $k$ is in units of $h$ Mpc$^{-1}$.

The transfer functions in figure 1 are bracketed by the transfer functions for CDM and HDM ($h = 0.5$ and $\Omega_B = 0.01$) found by Holtzman (1989). For comparison, in bottom panel of figure 1, we show his transfer functions for MDM models.

Perturbations on the largest scales enter the horizon after $t_{eq}$ and after the massive neutrinos have become non-relativistic. Growth on these scales is unimpeded, and the power today reflects directly the primordial spectrum. On smaller scales there are two effects. First, subhorizon-sized perturbations do not grow until $t_{eq}$. This explains the break in the CDM transfer function at $k \sim 0.1$ Mpc$^{-1}$. Second, relativistic particles can freestream out of dense regions and therefore subhorizon-sized perturbations in relativistic matter fields are severely diminished. As noted by Bond, Efstathiou, & Silk (1980), the freestreaming scale is

$$k_{FS} = \frac{2\pi}{\lambda_{FS}} = 0.5 \text{ Mpc}^{-1} \left(\frac{m_x}{100 \text{ eV}}\right). \qquad (21)$$

We see this in the fact that the scale at which the WDM curves first deviate from the CDM curve decreases in scale (i.e., increases in $k$) as we increase the mass to temperature ratio. The neutrinos in MDM are much lighter and hence freestream over much larger scales. This reduces the power spectrum at $k \gtrsim 0.1$Mpc$^{-1}$ [since MDM neutrinos constitute only a small fraction of matter, not *all* power is damped; CDM power remains].

We now show that the transfer function depends only on the velocity dispersion of the massive neutrino: $m_x/T_x \propto m_x/\alpha$. It is useful to carry out the computation of the transfer function in terms of the variable $q \equiv p/T_x = p/T_\gamma \alpha$. We see that $p/E = q/\sqrt{q^2 + (m_x a/T_\gamma \alpha)^2}$ depends only on $\alpha/m_x$. Moreover, we can change the integration variable in Eqs. (16) and (17) from $p$ to $q$. The integrals can then be written as an integral over $q$ which depends on $\alpha/m_x$ times $\alpha^4 \beta$. For example the integral in Eq. (16) becomes

$$\beta \alpha^4 T_\gamma^4 \int \frac{d^3 q}{(2\pi)^3} \, q \left[\sqrt{q^2 + \left(\frac{m_x a}{T_\gamma \alpha}\right)^2} + \frac{q^2}{\sqrt{q^2 + (m_x a/T_\gamma \alpha)^2}}\right] \frac{-\partial f_{0,x}}{\partial q} \Delta_x \; . \qquad (22)$$

Therefore, the only dependence on $m_x$, $\alpha$, $\beta$ is through the two combinations $\alpha/m_x$ and $\alpha^4 \beta$. But the latter is simply related to the former via Eq. (6). So we conclude that the power spectrum depends only on $\alpha/m_x$; in words, it depends only on the ratio of the heavy neutrino temperature to its mass.

## 4  Linear Tests

### 4.1  Fixing the mass

We want to determine the optimal value of the WDM mass. To do this, we focus on excess power ($EP$), a quantity which measures the relative mass excess on $25 \, h^{-1}$Mpc and $8 \, h^{-1}$Mpc scales (Wright *et al.* 1992). In addition, linear theory is used to estimate the epoch of galaxy formation. To facilitate these calculations, we use analytic fitting functions for the transfer functions found in the previous section. These are given in the Appendix.

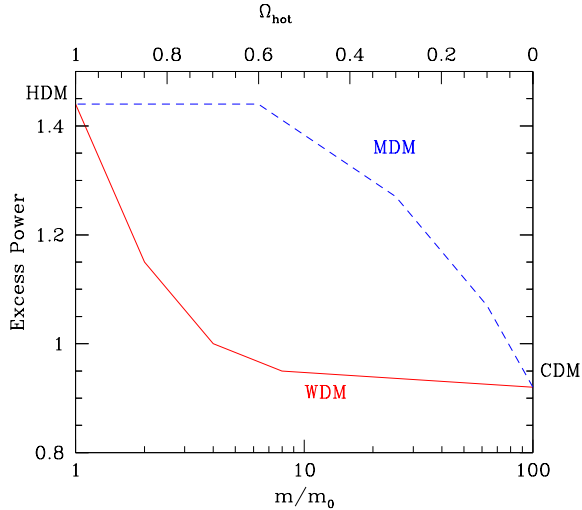

Figure 2: Excess power in theories interpolating between HDM and CDM. Solid curve shows how WDM with its free parameter $m$ [lower axis] interpolates. Note how it quickly becomes similar to CDM. Dashed curve shows the interpolation of MDM with its free parameter, the energy density in neutrinos [upper axis]. The observationally prefered valueof EP is 1.3.

It is generally accepted that the power in density fluctuations on $25\,h^{-1}$ Mpc relative to $8\,h^{-1}$ Mpc is greater in the data than in the standard CDM model. To quantify this, we first define the linear rms density fluctuations on a scale $R$:

$$\sigma_R \equiv \langle (\Delta M/M)^2 \rangle^{1/2} = \left( \int \frac{k^2 dk}{2\pi^2} P(k) W^2(kR) \right)^{1/2}, \qquad (23)$$

where $M \simeq 1.2 \times 10^{12} h^2 M_\odot \left( R/\mathrm{Mpc} \right)^3$ is the total mass in a sphere of radius $R$ and $W(x) = 3\left( \sin x - x\cos x \right)/x^3$ is the top hat window function. Wright *et al.* (1992) introduce the quantity EP defined as

$$EP \equiv 3.4 \frac{\sigma_{25}}{\sigma_8}. \qquad (24)$$

This definition is such that EP = 1 for standard CDM ($h = 0.5$, $\Omega_B = 0.1$, $\Omega = 1$) whereas consistency with the APM angular distribution function (to be discussed below) requires EP = $1.30 \pm 0.15$. Note that EP is independent of normalization, or equivalently biasing. The results for our family of WDM models are shown in Figure 2. For comparison, we also give EP in MDM as a function of the hot dark matter fraction. As expected, EP decreases as we increase the mass of the WDM particle. Our results for $m \gg m_0$ agree with those for an $\Omega_B = 0$ CDM model and we expect that like CDM, the EP calculated for WDM with a more realistic $\Omega_B = 0.05 - 0.1$ will be $5 - 10\%$ higher than in the $\Omega_B = 0$ case. With this in mind, we conclude that an $m \simeq 2m_0$ WDM model will have sufficient large-scale power to be in agreement with the APM results.

To go further we must normalize the power spectrum of Eq. (20). The COBE satellite (Smoot et al. 1992) has measured fluctuations in the cosmic microwave background on large angular scales where $T(k) \simeq 1$. These measurements are consistent with a spectral index $n = 1$ for the primordial perturbations. This is also the value predicted in the simplest models of inflation and is the value used in our analysis. (See, e.g., Adams *et al.* 1993 for a detailed discussion of cosmological models with different values of $n$.) Following Efstathiou, Bond, & White (1992) (more recently, see Bunn, Scott, & White, 1994; Gorski et al. 1994) we use the COBE results to determine the normalization constant of the power spectrum in Eq. (20):

$$B = \left(\frac{6\pi^2}{5}\right)\left(\frac{2}{H_0}\right)^4 \left(\frac{Q_{\rm rms-ps}}{T_0}\right)^2. \tag{25}$$

Here $T_0 = 2.726 \pm 0.006$ (Mather *et al.* 1994) is the present temperature of the microwave background. The first year COBE data gave $Q_{\rm rms-ps} = 17\mu K$; this is the normalization we have chosen for the N-Body runs described in section 5. Numerically, this gives $B = 6.0 \times 10^5 h^{-4}\,{\rm Mpc}^4$. The two year data have come in closer to $Q_{\rm rms-ps} = 20\mu K$ so we might be slightly underestimating the amplitude of the power-spectrum. A higher amplitude would however amplify, not alter, our conclusions.

Large-scale streaming velocities measure the mass fluctuations directly and can therefore be used to test and constrain models. For example, Bertschinger *et al.* (1990) estimate the three-dimensional velocity dispersions of optically selected galaxies within spheres of radius $40\,h^{-1}\,{\rm Mpc}$ and $60\,h^{-1}\,{\rm Mpc}$ and find $\sigma_v(40) = 388\,(1 \pm 0.017)\,{\rm km\,s}^{-1}$ and $\sigma_v(60) = 327\,(1 \pm 0.025)\,{\rm km\,s}^{-1}$. However, on such large scales, the power spectrum is independent of model type, at least within the class of models considered here, and therefore these measurements can only provide an alternative to COBE normalization. For the moment, the COBE measurements appear to be on firmer ground; streaming velocities are consistent with COBE but provide no additional constraints.

In the simplest models of galaxy formation, there is a single biasing parameter, $b$ such that $b\sigma_R$ gives the fluctuation in optically selected galaxies on the scale $R\,h^{-1}{\rm Mpc}$. Davis and Peebles (1983) find that $b\sigma_8 \simeq 1$ and therefore $1/\sigma_8$ is a measure of the optical bias. For WDM with $m = 2m_0$, $\sigma_8 = 1.0(Q_{\rm rms-ps}/17\mu{\rm K})$, significantly lower than the CDM value of 1.24 (recall that this is for low $\Omega_B$).

Perhaps the greatest difficulty with HDM is in forming galaxies at sufficiently early times. One way to estimate early galaxy formation is to calculate the mass excess on $0.5 h^{-1}$ Mpc scales, $\sigma_{0.5}$. This gives a rough (and probably low) estimate for the epoch at which structures on this scale went non-linear (see e.g. Bond & Efstathiou 1991 and Adams *et al.* 1993):

$$1 + z_{\rm gf} = \sigma_{0.5} \tag{26}$$

We find $\sigma_{0.5} = 1$, 1.7, 2.7, 3.8 for $m_x = m_0$, $2m_0$, $4m_0$, and $8m_0$ respectively. These results suggest that WDM will have trouble with early galaxy formation, a potential problem shared by MDM models with $\gtrsim 30\%$ of mass density in the hot component. Of course, galaxy formation necessarily involves nonlinear and non-gravitational physics (e.g. hydrodynamics) and so these conclusions should be used with caution.

## 4.2 Linear Power Spectrum Versus Observations

With the "best fit" mass for WDM now set at $m = 2m_0$, we can compare the full power spectrum to the data. Recently, Peacock & Dodds (1994, PD) attempted to reconstruct the linear power spectrum $P_{\rm L}^{\rm g}(k)$ of the underlying matter distribution from the observed galaxy distribution. They assumed a simple linear relationship between the matter power-spectrum and the galaxy power-spectrum and in addition, corrected for redshift distortions and nonlinear dynamics. The results of PD are displayed in figure 3, with some modifications: we do not show the very small errors computed by PD but instead use errorbars based on a simple visual estimate of the vertical scatter in their figure 6. Furthermore, in the left panel of figure 3, we multiply the amplitude of their $P_L^g(k)$ by a factor $b^2 = 1.3^2$ to normalize it to the optical galaxy distribution. In the right panel, we keep their normalization to IRAS galaxies (Strauss et al. 1992).

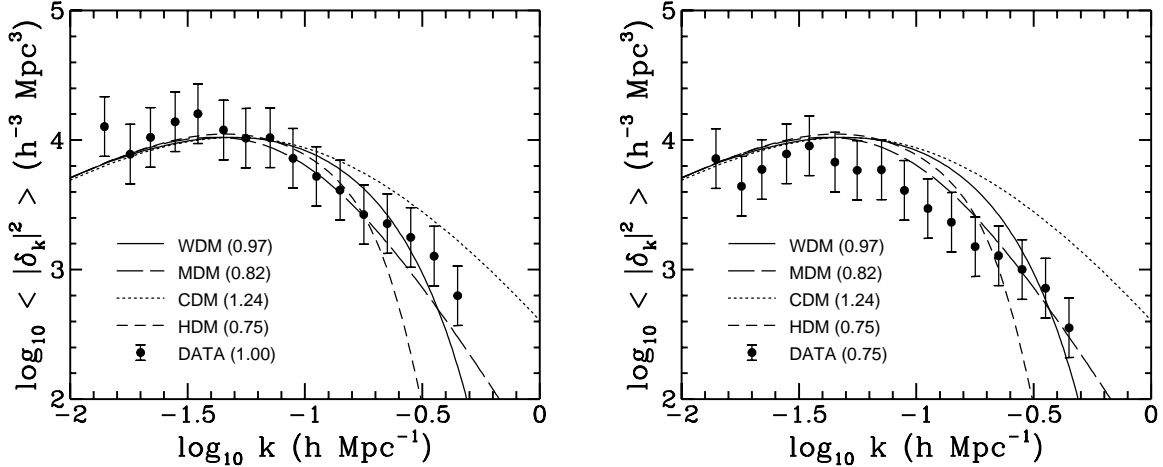

Figure 3: Linear power spectra of the WDM, MDM, CDM and HDM distributions, compared to the observational data compiled by Peacock & Dodds (1994, PD). In left panel, the measurement of PD has been enhanced by a factor $1.3^2$ to match the optical galaxy normalization ($\sigma_8 = 1$). The right panel is the same as the left one, but the dots are normalized to IRAS galaxies ($\sigma_8 = 0.75$). The errorbars we put on the dots are also much larger than those quoted by PD (see text). The number in parentheses gives the value of $\sigma_8$ for the considered power-spectrum.

Figure 3 also shows the predicted power spectra for four models: CDM, WDM with $m = 2m_0$, MDM with $\Omega_{\rm hot} = 0.3$, and HDM. All the spectra assume $\Omega_B = 0.01$ except for WDM which has $\Omega_B = 0$. CDM, MDM and HDM spectra are extracted from Holtzman (1989). The number in parentheses gives the linear value of $\sigma_8$, with our assumed value of $Q_{\rm rms-ps} = 17\mu K$.

COBE normalization, together with the assumption that the IRAS galaxy distribution closely follows the underlying matter distribution, appears to be incompatible with CDM, HDM, possibly WDM and only marginally compatible with MDM. The situation improves if we normalize instead to optical galaxies. In any case, all of the models have the same power spectrum for $k \lesssim 0.1 h {\rm Mpc}^{-1}$ and can be distinguished from one another only for $k \gtrsim 0.1 h {\rm Mpc}^{-1}$. The best fit to the data seems to be MDM. WDM is not too bad, although it has a bit too much power at intermediate scales $-1.2 \lesssim \log_{10} k \lesssim -0.7$, particularly if the comparison of the power-spectrum is made with data normalized to IRAS galaxies. CDM has of course too much power at small scales and HDM not enough.

## 5  $N$-Body Experiments

This section discusses the results of our WDM, CDM and HDM $N$-body experiments. § 5.1, outlines the simulations. We make a visual analysis in § 5.2 comparing redshift "slices" of HDM, CDM and WDM "galaxy" distributions with the CfA2 slice of de Lapparent et al. (1986). In § 5.3, we analyse the pairwise statistical properties of the matter distribution, such as the power-spectrum, correlation function and line-of-sight velocities, and compare the results to observations. Section 5.4 discusses the cluster multiplicity function.

## 5.1 The Simulations

We now discuss the results of N-body simulations for WDM ($m_x = 46$ eV; $T_x = (4/11)^{1/3}$), HDM, and CDM. Five simulations, four with the Particle-Mesh (PM) code of Moutarde *et al.* (1991) and one with the treecode (TREE) of Bouchet & Hernquist (1988, later improved by Hernquist, Bouchet & Suto, 1991) are run for each of the models. For the PM simulations, a $128^3$ grid is used to compute the forces with either $64^3$ or $128^3$ particles. The TREE simulations involve $32^3$ particles and are used primarily to check the accuracy of the PM simulations at small scales. The very large scale regime is probed by PM simulations with $128^3$ particles and a physical box size $L_{\text{box}} = 720$ Mpc. In these simulations, the mass of each particle is rather large ($M_{\text{part}} = 1.23 \times 10^{13} M_\odot$). The physical size of the other simulations (hereafter PMS, PMS64a, PMS64b and TREE) is $L_{\text{box}} = 144$ Mpc, with a corresponding particle mass $M_{\text{part}} = 9.88 \times 10^{11} M_\odot (128^3/N_{\text{par}})$ which is about the mass of a galaxy for $N_{\text{par}} = 128^3$. Table 1 summarizes the various parameters associated with each simulation.

Our models assume $h = 0.5$, $\Omega_B = 0$, and $\Lambda = 0$. Initial conditions (scale factor $a \equiv 1$) are generated from the linear power spectrum by slightly perturbing a regular pattern of particles using the Zel'dovich approximation (Zel'dovich, 1970). The amplitude of the initial fluctuations is set so that the density fluctuations on 16 Mpc scales is $\sigma_8 = 1/16 = 0.0625$ ($\sigma_8 = 1/8 = 0.125$ for the TREE simulations). The simulations are then evolved until the linear power spectrum reaches the COBE normalization ($Q_{\text{rms-ps}} \simeq 17$) corresponding to a final scale factor $a = 20$, 16, 12 respectively for CDM, WDM and HDM ($a = 10$, 8, 6 for the TREE simulations). Although we studied several stages of the simulations, we analyze here only the last snapshot. We have neglected possible free-streaming effects for WDM and HDM: they should be very small during the period covered by our simulations at the scales we are interested here ($\gtrsim 1$ Mpc).

We have checked that the measurements of the two-body correlation function and the line-of-sight velocity dispersion (defined in § 5.3.3) for our CDM simulations are in reasonable agreement with those of Gelb & Bertschinger (1994) and Zurek *et al.* (1994), who did high resolution CDM simulations with large numbers of particles. We have not yet compared the results of our WDM and HDM simulations to large high resolution simulations (this is left for future work). We may in fact be underestimating the small scale velocities dispersions (§ 5.3.3), though a preliminary comparison of a TREE simulation to a PM simulation, both starting from the same initial conditions and using $64^3$ particles, suggests that the discrepancy will be less than 30%. The discrepancy between high and low resolution codes should be much less pronounced for analysis of the statistical properties of the density distribution.

## 5.2 Visual impression

Figure 4 displays thin ($L_{\text{box}}/64$ thick) slices of the simulations PML and PMS. The panels from top to bottom correspond to CDM, WDM and HDM. Figure 5 is the same, but the slices are thicker ($L_{\text{box}}/32$ in the left panels and $L_{\text{box}}/4$ in the right ones) and only overdense regions are kept. These regions are found using one of the following two methods:

(i) For the large PML simulations (left panels), we assume that galaxies form in weakly evolved overdense regions. We take this epoch of "galaxy formation" to be when $a = 2$ corresponding to a redshift $z_{\text{gf}} = 9$, 7, and 5 for CDM, WDM and HDM respectively. At this scale factor, we select particles that have at least one neighbour closer than $A = 0.95$ times the mean interparticle distance $d$ and follow them until the present time. This procedure amounts to selecting overdense regions bounded by isosurfaces with densities at $z_{\text{gf}}$ of order $\rho/\overline{\rho} \sim 2/A^3 \sim 2.33$. The corresponding density contrast at the present epoch (if one naively applies linear theory) is $\delta\rho/\overline{\rho} \sim 27$, 20, 16 for CDM, WDM and HDM respectively.

Table 1: Characteristics of the simulations

| Name | $\lambda_{\min}(\text{Mpc})$ [a] | $N_{\text{part}}$ [b] | $L_{\text{box}}(\text{Mpc})$ [c] |
|------|---------|---------|---------|
| PML | 5.625 | $128^3$ | 720 |
| PMS | 1.125 | $128^3$ | 144 |
| PMS64a | 1.125 | $64^3$ | 144 |
| PMS64b | 1.125 | $64^3$ | 144 |
| TREE | 0.225 | $32^3$ | 144 |

[a] spatial resolution. For the PM code, this scale corresponds to the size of a grid cell. For the treecode, this scale corresponds to the short range softening parameter $\epsilon$.
[b] mass resolution (number of matter particles).
[c] simulation box size.

Figure 4: Thin slices $L_{\text{box}}/32$ thick extracted from the simulations PML (lefts panels) of physical size $L_{\text{box}} = 720$ Mpc and the simulations PMS (right panels) of physical size $L_{\text{box}} = 144$ Mpc. The top, middle and bottom panels correspond respectively to the CDM, WDM and HDM model. *Figure 4 available directly from authors or via www at http://fnas08.fnal.gov/pub/Publications/Pub-95-093-A.*

(ii) For the PMS simulations (right panels), we consider the present epoch and use the friends-of-friends algorithm of Efstathiou *et al.* (1988, hereafter EFWD) to select connected groups of particles in which each element has at least one neighbour closer than $A = 0.2$ times the mean interparticle distance. These groups define regions of density larger than $\rho/\overline{\rho} \sim 2/A^3 \sim 250$. They are displayed in the right panels of figure 5 and will be used later to study the cluster multiplicity function.

A useful exercise is to make a direct comparison with the CfA redshift survey (de Lapparent, Geller & Huchra 1986; Geller & Huchra 1989). CfA-like slices are extracted from the catalogs of points displayed in the left panels of figure 5 and displayed in figure 6 along with the observed galaxy distribution (de Lapparent *et al.* 1986). The observer is assumed to be at the bottom of each slice. The slices have a depth of 12,800 km/s in redshift space, or 256 Mpc with our choice of $H_0$. The synthetic slices account for redshift distortions induced by the peculiar velocities of the galaxies. In addition, we model selection effects as follows: given the magnitude limit 15.5 of the CfA survey and the Schecter form (Schechter 1976) for the galaxy luminosity function (with parameters measured by de Lapparent *et al.*, 1991) we compute the average number density $n_D$ of selected galaxies in a thin shell at a distance $D$ from the observer. The probability that a matter particle at a distance $D$ is included in the synthetic survey is then $n_D/n_S$ where $n_S$ is the average number density of "galaxies" in the $N$-body sample. When $D$ is small, we can have $n_D > n_S$ indicating that we undersample the real galaxy distribution. The contour $n_D = n_S$ is indicated by a dashed line on each figure.

Figure 5: Same as in figure 4, but only overdense regions, where galaxies are expected to remain, have been kept and the slices are thicker. In the left panels, the slices are $L_{\text{box}}/32$ thick; the matter particles belonging to regions of density larger than $\rho/\overline{\rho} \simeq 2.33$ have been selected at a weakly evolved stage $a = 2$ and followed until present time. In the right panels, the slices are $L_{\text{box}}/4$ thick; each point represent a connected group of particles belonging to regions of density larger than $\rho/\overline{\rho} \simeq 250$. *Figure 5 available directly from authors or via www at http://fnas08.fnal.gov/pub/Publications/Pub-95-093-A.*

Figure 6: CfA-like redshift slices of the observed galaxy distribution. The observer is located at the bottom of each slice. The slices are 256 Mpc deep (with our choice of the Hubble constant). The top panel represents a slice of the real observed galaxy distribution (courtesy V. de Lapparent). The others panels correspond to artificial catalogs built from the simulations, taking into account redshift-space distortions and selection effects (see text). From top to bottom, one passes from CDM to WDM and HDM. The central slices have the same geometry as the CfA slice, i.e. are covering the declination range $26.5° < \delta < 32.5°$. The left slices are the adjacent slices with $20.5° < \delta < 26.5°$ and the right ones are the adjacent slices with $26.5° < \delta < 32.5°$. All the slices are projected on the plane $\delta = 0$, and rescaled so that they have all the same apparent size. The small dotted arcs of a circle determine a limit below which we undersample the observed galaxy distribution (see text).*Figure 6 available directly from authors or via www at http://fnas08.fnal.gov/pub/Publications/Pub-95-093-A.*

To facilitate comparisons between WDM, CDM and HDM, we use the same random numbers to set the initial conditions for each simulation. By construction, the power spectra have the same normalization at the COBE scale and therefore each model should present similar features at very large scales. This is indeed the case. The WDM model considered here is, as expected, pancake-like rather than hierarchical with a smooth density distribution similar to the one found in the HDM simulations. However, as in the CDM case, the WDM distribution exhibits rich, dense and almost spherical clusters which are certainly virialized. Such clusters are absent, or at best very rare, in our HDM simulations. Indeed the dense regions in the HDM simulations are still sheet-like or filamentary, i.e., not yet virialized. One can also see (right panels of figure 5) that the WDM distribution presents nice large filamentary structures. This is also the case in the CDM distribution (e.g. West, Villumsen & Dekel 1991), but there the filaments tend to be broken into clumpy substructures.

We also see from figure 5 that the apparent size of the underdense regions or voids increases as one passes from CDM to WDM and HDM, in agreement with earlier studies (e.g., Melott 1987). HDM appears to be ruled out because the voids are too large as compared with the CfA data (Zeng & White 1990). The voids in the WDM simulation are still a bit too large. On the other hand, CDM nicely reproduces the qualitative features of the CfA slice, as already stated by White *et al.* (1987).

The dense structure in the center of the CfA slice corresponds to the Coma cluster. The fact that it is elongated is due to the high internal velocity dispersion of this cluster. We do not have such strong effects in our synthetic slices, not because our models do not produce such clusters (we shall see later that on the contrary, the small scale velocity dispersions are quite large), but because our PM code tends to underestimate small scale velocities. Indeed, the resolution of the simulations used to build the slices is about 6 Mpc, which is typically the size of a rich cluster.

## 5.3 Matter distribution properties

This section is devoted to the pairwise properties of the matter distribution. In particular, we consider the evolved power-spectrum $P(k) \equiv \langle |\delta_k|^2 \rangle$ (§5.3.1), the two-point correlation function (§5.3.2), and the pairwise velocity dispersion (§5.3.3). When comparing with the data, we must remember that the simulations give information only on the mass distribution, while observations probe the distribution of galaxies. In §5.3.4, we discuss briefly how the difference between the two – so-called biasing – influences our interpretation. Our basic conclusion is that non-linear effects substantially tarnish the optimistic view we gained in section 4, when we included only linear effects.

### 5.3.1 Power spectrum

Figure 7 shows $P(k)$ for WDM, CDM and HDM. For each simulation, we compute the density field $\rho(x)$ in a grid of resolution $128^3$ using a Cloud-In-Cell (CIC) scheme (see, e.g., Hockney & Eastwood 1981). The power spectrum is then obtained by fast fourier transform. The calculation is done for $2\pi/L_{\rm box} \lesssim k \lesssim k_{\rm ny}/3.2$ where the results are only weakly contaminated by nonphysical, numerical effects, such as white noise or the smoothing introduced by the CIC affectation. Here $k_{\rm ny}$ is the Nyquist frequency of the grid used to compute the power spectrum. The curves represent averages over all simulations (including TREE) and the errorbars correspond to the rms dispersion.

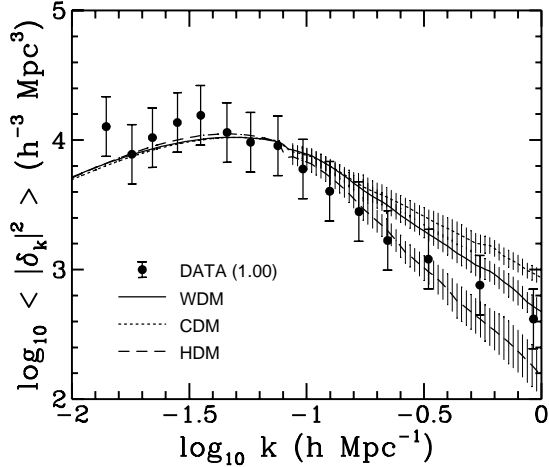

Figure 7: Power spectrum measured in the WDM, CDM and HDM simulations. For $\log_{10} k \leq 1.1 h^{-1}$Mpc, linear theory is used (non-linear effects are negligible on such scales). The dots correspond to the data used by Peacock & Dodds (1994, PD), enhanced by a factor $1.3^2$ to match the optical galaxy power-spectrum. The errorbars on the dots are our own and are much larger those quoted by these autors (see § 4.2).

The nonlinear power spectra are much closer to each other than are the linear ones. In particular, it is difficult to distinguish WDM from CDM. This is not so surprising: as already noticed for example by EFWD, an expanding collisionless medium subject to gravitational instability seems to evolve towards self-similar behavior that is only weakly dependent on initial conditions. Essentially, power cascades down from large scales to small scales as the system enters the nonlinear regime. Since the initial power spectra considered here have roughly the same shape at large scales the differences between CDM, WDM and HDM tend to decrease with time as the system relaxes. Our first important conclusion then is that *non-linear effects make the power-spectrum of WDM look very much like CDM*.

The data points in figure 7 correspond to the nonlinear power-spectrum $P^{\rm g}_{\rm NL}(k)$ infered from $P^{\rm g}_{\rm L}(k)$ using the mapping of PD. In other words, to be able to compare our nonlinear power-spectra to their measuments, we omit the step in their calculation which consists of going back in time to obtain the linear power-spectrum. In principle, figure 7 should lead to the same conclusions found in figure 3 (left panel) where we used the linear power spectra, $P^{\rm g}_{\rm L}$. This is approximately true for CDM, but not quite for WDM and HDM, particularly at the smallest scales shown in figure 7. However, this is not very surprising since the mapping of PD is expected to be less accurate for pancake models. We therefore expect the non-linear comparison in figure 7 to give the more realistic comparison between our models and the measurements.

Even with our generous errorbars, the CDM distribution has too much power at small scales confirming earlier findings. WDM, like CDM, seems to systematically overestimate the observations for $\log_{10} k \gtrsim -1.0$, particularly around $\log_{10} k = -1$. The HDM distribution provides a very good fit at large scales ($\log_{10} k \lesssim -0.7$) but with too little power on small scales. Biasing, as will be discussed in § 5.3.4, or normalization of the data point to IRAS galaxies (see § 4.2), probably

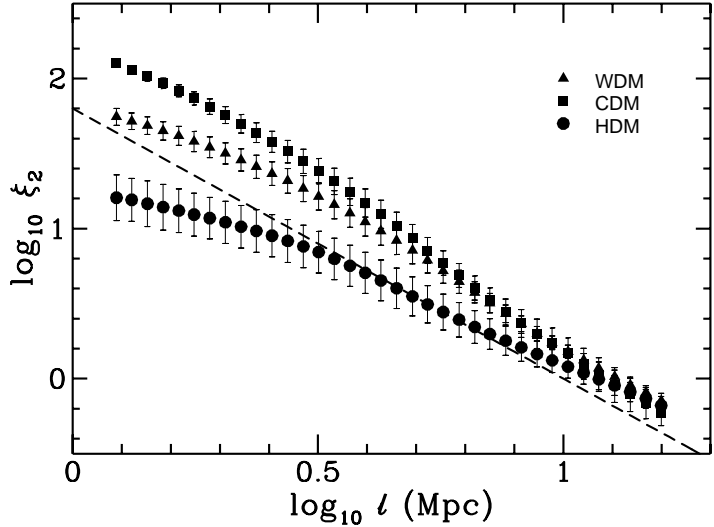

Figure 8: Measured two body correlation function in our WDM, CDM and HDM simulations. The dashed line is the power-law fit of Davis & Peebles (1983) in the observed galaxy distribution.

worsens the situation for WDM.

### 5.3.2 Correlation function

Figure 8 displays the two-body correlation function $\xi_2(\ell) \equiv \langle \delta(\mathbf{x})\delta(\mathbf{x}+\ell)\rangle$ where $\delta \equiv \delta\rho/\overline{\rho}$ is the density contrast. Since the two-body correlation function is just the fourier transform of the power spectrum (see, e.g., Peebles 1980), we expect similar conclusions. For each simulation (except PML), we measure $\xi_2(\ell)$ and average the results. The analysis is done for $L_{\rm box}/128 \leq \ell \leq L_{\rm box}/9$, where the lower bound corresponds to the spatial resolution of the PM code and the upper one is imposed to avoid possible contamination due the finite size of the simulation box. The errorbars represent the rms dispersion of the simulations. The dashed line is the power-law fit $\xi_2^{\rm G}(\ell) = (\ell/10.8)^{-1.77}$ of the two-body correlation function measured by Davis & Peebles (1983) in the optical galaxy distribution.

As expected, the results are similar to those of § 5.3.1. In particular, the function $\xi_2$ measured in the WDM distribution is very close to the one measured in the CDM distribution, although its overall logarithmic slope is closer to the observed one. In both cases, the measurements overestimate by a significant amount the optical correlation function and therefore require some "antibias" between the galaxy distribution and the matter distribution, i.e. $\xi_2^{\rm G}(\ell) = b^2(\ell)\xi_2(\ell)$, with $b(\ell) < 1$. For example, at the correlation length of the optical galaxy distribution $\ell_0 \simeq 10.8$ Mpc, we measure $b(\ell_0) = 0.8$ for CDM and WDM, and $b(\ell_0) = 0.9$ for HDM. We return to this point in § 5.3.4.

### 5.3.3 Pairwise velocities

The line-of-sight pairwise velocity dispersion

$$\sigma_1(r) \equiv \frac{1}{\sqrt{3}} \left\langle [\mathbf{v}(\mathbf{x}+\mathbf{r}) - \mathbf{v}(\mathbf{x})]^2 \right\rangle^{1/2} \qquad (27)$$

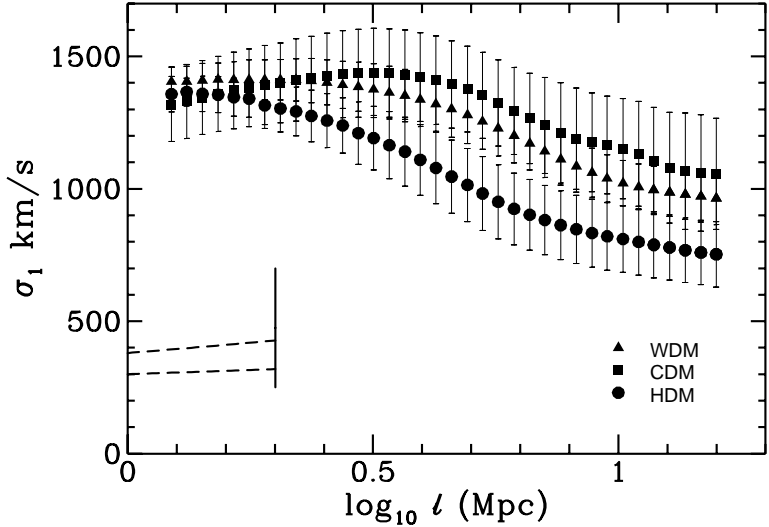

Figure 9: The quantity $\sigma_1$ (defined in Eq. [27]) as function of separation $\ell$ measured in our $N$-body experiments, compared to measurements in the observed galaxy distribution. The two dashed lines at the bottom left correspond to the measurement of Davis & Peebles (1983) on the CfA1 catalog and the thick vertical segment to a compilation of the more recent measurements of Mo *et al.* (1993) on various galaxy catalogs (see text).

provides another probe of structure on galaxy and cluster scales. Here, $\mathbf{v}(\mathbf{x})$ stands for the peculiar velocity of the matter measured in our simulations. $\sigma_1(r)$ calculated in the synthetic data can be compared (with caution) to measurements in the galaxy distribution as is done in figure 9. The analysis for the simulations is similar to the one used to calculate the two-body correlation function. The errorbars, which represent the rms dispersion over all of the synthetic data sets, are quite large especially for $r < 10$ Mpc where $\sigma_1$ is dominated by rare, large, and hot (high internal velocity dispersion) clusters (see also GB).

Once again, we see that the results for WDM and CDM are fairly close. Even HDM gives similar results at small scales. This last point apparently contradicts the results of the previous sections, which looked at the statistics of the density distribution. However, the collapse of large cluster-like objects can produce large velocity dispersions at small separations, particularly just after the first shell crossing (see also Gelb, Gradwohl, & Frieman 1993).

The dashed lines at the bottom left of figure 9 bracket the observed values of the line-of-sight pairwise velocity dispersion $\sigma_1^g$ (Davis & Peebles 1983) as measured in the CfA1 galaxy catalog (Huchra *et al.* 1983). The thick vertical line corresponds to a compilation of more recent measurements made by Mo, Jing & Börner (1993) using both the CfA1 and CfA2 catalogs (Huchra *et al.* 1990) as well as the SSRS catalog (da Costa *et al.* 1991) and the 1.930 Jy redshift survey of IRAS galaxies (Strauss *et al.* 1992). For separations $0.8h^{-1}$ Mpc $\lesssim r \lesssim 1.6h^{-1}$ Mpc, we have 280 km/s $\lesssim \sigma_1^g \lesssim$ 700 km/s with the one exception being the result for the CfA2 catalog that we did not take into account while drawing the vertical line. Indeed, this last catalog is dominated by the coma cluster $\sigma_1^g \sim 1400$ km/s, a value close to the one we measure in our $N$-body experiments. Except for this particular measurement, the observed $\sigma_1^g$ is significantly less than the $\sigma_1$ found in the simulations.

The models exhibit small-scales velocities more than a factor two larger than those observed in the galaxy distribution suggesting that they are excluded by the data. However, there are still large uncertainties in the measurements (Zurek *et al.* 1994; Mo *et al.* 1993). In addition, there is the usual problem that the velocity dispersion measured for galaxies may be different from the velocity dispersion for the underlying matter distribution. We now turn to this everpresent question of biasing.

### 5.3.4 Biasing

The preceding subsections have all illustrated that non-linear effects substantially enhance the power in a WDM model at scales $k \gtrsim 0.1 h \text{Mpc}^{-1}$. The observations of galaxy distributions seem to indicate that there is *less* power on these intermediate scales than the model predicts. One way to reconcile this discrepency would be to invoke "anti-biasing," i.e. assume that $P^{\text{g}}/P$ and $\sigma_1^{\text{g}}/\sigma_1$ are less than one. There are two problems with this solution. First, the extensive studies of biasing in CDM models suggest that the biasing parameter $b \equiv (P^{\text{g}}/P)^{1/2}$ is larger than unity. One might argue that WDM may be biased differently since it is not a "hierarchical" model like CDM. This leads to the second problem: there have been some studies of biasing in pancake models and these suggest that the bias factor $b$ is *larger* than in hierarchical models. The situation for velocities is slightly better. The velocity bias parameter defined here as $b_{\text{v}} \equiv \sigma_1^{\text{g}}/\sigma_1$ is expected to be less than unity for it is difficult to imagine a mechanism which can accelerate the baryonic matter but not the dark matter. Both merging (Couchman & Carlberg 1992) and dynamic friction inside clusters (Carlberg & Dubinski 1991) may significantly decelerate the galaxies relative to the dark matter thereby leading to a low $b_{\text{v}}$.

We first review the work on biasing. Perhaps the simplest method (i) is to assume that galaxies form in regions with densities larger than a given threshold and that their distribution follows the matter distribution in these regions (e.g. Einasto, Klypin, & Saar 1986 and references therein). This is basically the method used to generate the left panels of figure 5, though there the "galaxies" were selected at some reasonable epoch of galaxy formation and then followed until the present. A more elaborate approach (ii) is to assume that galaxies form in the peaks of the matter distribution (see, e.g., Davis *et al.* 1985; Bardeen *et al.* 1986). These two methods lead to values of $b$ larger than unity (at least for gaussian initial fluctuations). Another procedure (iii), which makes use of a friends-of-friends algorithm to select connected groups of particles to identify halos of galaxies (Frenk *et al.* 1988), can lead to antibias $b < 1$, particularly at small scales. However this result depends strongly on the way large halos are treated. If large halos have significant substructure and correspond to several galaxies rather than only one then the bias will be larger and probably greater than unity (Gelb & Bertschinger 1994, hereafter GB). Further refinements can be added to the above recipies (see for example White *et al.* 1987; Klypin *et al.* 1993 and Nolthenius *et al.* 1994; Gelb & Bertschinger 1992; Carlberg 1988, 1991; Fry & Gaztañaga 1993). In addition, one can attempt to treat the collisional nature of the luminous matter (e.g., Katz, Hernquist & Weinberg, 1992, Cen & Ostriker, 1992). In general, one finds that $b$ is larger than unity. The bias is however deeply related to the merging history of galaxies and to the way galaxies form in clusters: values of $b$ smaller than one are still not excluded for CDM (see, e.g., Couchman & Carlberg 1992, Zurek *et al.* 1994).

Velocity bias has been studied in detail for the CDM model, but there is no real agreement yet in the scientific community. Current estimates indicate $0.5 \lesssim b_{\text{v}} \lesssim 1$ (e.g., Couchman & Carlberg 1992; Cen & Ostriker 1992; Katz, Hernquist & Weinberg 1992; Carlberg 1994, Zurek *et al.* 1994,

GB). The first two methods (i) and (ii) of galaxy selection invoked above, which assume that galaxies form in overdense regions or in the peaks of the density distribution, lead to a velocity bias only slightly smaller than unity. Friends-of-friends algorithms (iii) can lead to a significant velocity bias of order $b_v \sim 0.5$ or even smaller. Indeed, the selected objects can be rich halos with high internal velocity dispersions whereas $\sigma_1$ takes into account only the average (barycentric) velocity. If however very massive halos fragment into smaller components, (i.e., correspond to several galaxies instead of just one) the velocity bias would be larger and probably close to unity (GB).

It is not obvious how to implement biasing in a pancake model. The difficulty is that the matter is organized in thin sheets and so it is difficult to identify halos. The most naive approach (i) is to assume that galaxies form in the overdense parts of the matter distribution. This leads to the second left panel of figure 5. The power spectrum of this WDM distribution is approximately *twice as large* as the one directly measured in the full WDM distribution. The idea that the power spectrum is strongly enhanced in WDM if galaxies form in the overdense parts of the matter distribution agrees with earlier studies of HDM (White, Frenk, & Davis 1983; Braun, Dekel & Shapiro, 1988). Of course, processes of galaxy formation are not simple, and one can find arguments that reduce such an enhacement, such as the feedback from the first generation of formed objects in the luminous distribution (Braun, Dekel & Shapiro 1988). Recent analyses of the HDM model, including the hydrodynamics of the gaseous component (Cen & Ostriker 1992) seem however to confirm the above simple view that the galaxy power-spectrum is larger than the matter power-spectrum in pancake models.

To summarize, with the current observational data, the models we are studying require $b < 1$, $b_v \lesssim 0.5$. While certainly not impossible, this seems rather unlikely.

## 5.4 Group multiplicity function

The multiplicity function (Gott & Turner 1979), essentially the density of groups and clusters as a function of the number of objects they contain, can be quite useful in testing structure formation scenarios. Following Weinberg & Cole (1992), we measure the multiplicity function in our $N$-body experiments and compare the results to those of Moore, Frenk & White (1993, hereafter MFW) for the CfA galaxy catalog.

By definition, a group of particles in our synthetic data will have the multiplicity $X$ if it involves $N$ members with $2^{X-1} < N \leq 2^X$. The multiplicity function $n(X)$ is then the number density of groups with multiplicity $X$. The groups themselves are selected with the friends-of-friends algorithm of EFWD and are thus connected sets of particles for which each member has at least one neighbour closer than $A = 0.2$ times the mean interparticle distance. Right panels of figure 5 display the groups selected in this way from our PMS $N$-body simulation.

The measurement of the multiplicity function in the observed galaxy distribution is quite a delicate matter. Indeed, in three dimensional galaxy catalogs, the apparent number density of galaxies decreases with distance due to selection effects. In addition, peculiar velocities of galaxies distort estimates of their distances. MFW correct for these effects and derive a luminosity function $\widetilde{n}(L)$ of groups. To do this, they used a friends-of-friends algorithm similar to the one of EFWD but modified in order to take into account observational effects (see also Huchra & Geller 1982, Geller & Huchra 1983, Nolthenius & White 1987). We use the measurements of MFW for groups with similar overdensity to the one of our groups $\delta\rho/\overline{\rho} \sim 250$ ($D_0 = 1.0$ Mpc in their notations, see their Table 2). In order to convert their luminosity function to a multiplicity function, one must make some assumptions about the mass to light ratio for the groups. The simplest assumption is that $M/L$ is the same for all objects. We use $M/L = 123\ M_\odot/L_\odot$ as estimated by MFW.

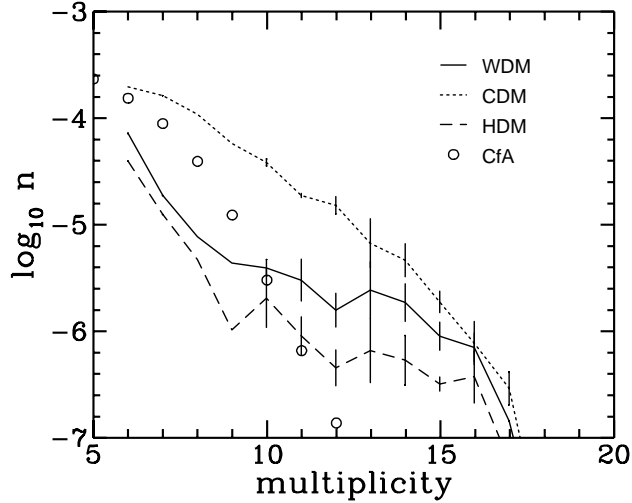

Figure 10: The multiplicity function measured in our $N$-body experiments (curves with errorbars, see text) compared to the measurement in the CfA galaxy catalog by Moore, Frenk & White (1993) (circles). We assume here that groups and clusters have a constant mass to light ratio $M/L = 123 M_\odot/L_\odot$, where $M_\odot/L_\odot$ is the mass to light ratio of the sun. The unit of mass choosen to compute the multiplicity is $M = 2.88\ 10^{11}$ solar masses. A cluster of mulplicity $X$ has a mass comprised between $2^{X-1}M$ and $2^X M$.

For the simulations, we assume that each matter particle corresponds to one member (i.e., galaxy) in a group. We determine $n(X)$ in each simulation (except TREE) and average the results. The analysis is made for $X \geq 6$ as the assumption that the mass to light ratio is the same for all group members may work only for groups with large numbers of objects. Figure 10 shows $n(X)$ for both the $N$-body simulations and the data. The error bars represent the rms dispersion between all the measurements. No errorbar indicates that there was only one measurement available.

The multiplicity function for WDM is closer to HDM than CDM, an indication that structure formation begins with the formation of large pancake-like objects. None of the models agree with the data at large multiplicity, at least for the mass to light ratio we choose. We can choose a different $M/L$ but this does not really help. In particular, $n(X)$ for WDM and HDM have the wrong shape and the one for CDM is not much better. In figure 11 we plot the mass to light ratio as a function of multiplicity required if the $N$-body results are to agree with the data. For $X \lesssim 10$, the $M/L$ required by CDM is comparable to the observed $M/L = 150 \pm 50 M_\odot/L_\odot$ (see, e.g., Peebles 1992) whereas the $M/L$ required by WDM and HDM are too small. At larger $X$, the required ratio $M/L$ increases with $X$ and becomes unrealistically large for all the models.

The above analysis indicates that the WDM, CDM and HDM models considered here all produce too many rich clusters. Moreover, WDM and HDM clearly exhibit the wrong shape for $n(X)$ provided one assumes that the mass to light ratio of clusters is constant or only weakly varying with richness, as is currently suggested by observations (see also Weinberg & Cole 1992). However, our analysis is rather crude and needs to be improved before making any final conclusions.

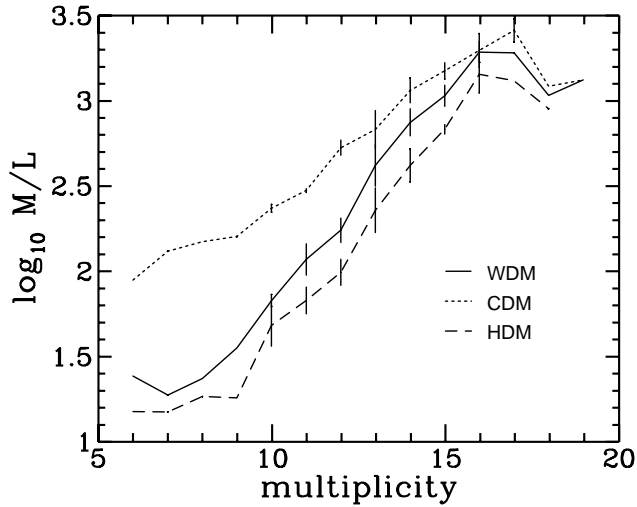

Figure 11: The mass to light ratio (in units of the mass to light ratio of the sun) that would be required for the multiplicity function measured in our $N$-body experiments to fit the one measured by Moore, Frenk & White (1993) on the CfA catalog (see Fig. 10).

## 6  Conclusion

Warm dark matter is an interesting and viable alternative to the standard CDM and HDM cosmologies. Quite generally, WDM refers to any particle whose velocity dispersion is non-negligible (for the purposes of structure formation) but less than the velocity dispersion for the standard HDM neutrino. We have studied a one-parameter family of WDM models where the distribution function for the dark matter candidate is given by Eq. (1). Here, we summarize our results.

1. By definition, $m = m_0$ corresponds to HDM. As $m$ is increased, the linear transfer function approaches that of CDM in a way that is qualitatively different from MDM models.

2. Linear analysis suggests that the $m = 2m_0$ WDM model satisfies observational tests which probe structure on scales greater than 25 $h^{-1}$ Mpc. These tests include EP (excess power on 25 $h^{-1}$ Mpc as compared with 8 $h^{-1}$ Mpc) and bulk velocities on 40 $h^{-1}$ Mpc − 60 $h^{-1}$ Mpc. In addition, the COBE normalized linear power spectrum provides a better fit to the data than either HDM or CDM. However, WDM may have problems in forming galaxies at sufficiently early times.

3. Detailed N-body simulations for CDM, HDM and WDM ($m = 2m_0$) are used to compare the models in the non-linear regime. As one might expect, WDM has properties of both HDM and CDM. In particular:

    - Structure formation in the WDM model studied is pancake-like rather than hierarchical. The density distribution is rather smooth and structures as small as galaxy halos are excluded.
      Rich, dense, almost spherical, and certainly virialized clusters appear. These are evident in the CDM simulations but not in the HDM simulations.

Simple visual analyses of the large scale structures such as filaments, sheets and large void suggest that WDM reproduces well the observed ones, although the voids may be slightly too large, but still significantly smaller than in HDM.

- The pairwise statistical properties of the WDM distribution look pretty much like in CDM (power-spectrum, correlation function, line-of-sight velocity dispersion). It thus presents more "power" at small scale than observations, implying an *antibias* $b < 1$ between the galaxy distribution and the matter distribution.

    All models predict velocities on small scales that are much higher than the velocities measured in the data though there are a number of both theoretical and observational uncertainties which could explain this discrepancy.

- The group multiplicity function, which estimates the density of groups or clusters of galaxies as a function of the number of objects they contain, is calculated for the three models and compared with the multiplicity function for the CfA galaxy catalog derived by Moore *et al.* (1993). The multiplicity function for WDM is similar to that of HDM illustrating the pancake-like nature of gravitational collapse in a WDM universe. Neither the HDM or WDM multiplicity functions have a shape in agreement with the data. CDM is not much better.

The primary purpose of this paper has been to see how the velocity space distribution function of the dark matter affects the formation of structure. We have therefore made a number of simplifying assumptions which allow for easy comparisons among the models. In particular, we set $h = 0.5$, $\Omega_B = 0$, $\Lambda = 0$ and assumed a simple form for the primordial perturbation spectrum. Our tentative conclusions are that within this context, warm dark matter does not agree well with the data. By varying these assumptions/parameters, however, WDM could do better. The results enumerated above may help discover a more fitting context for warm dark matter.

**Aknowledgements:** We thank B. Gradwohl, A. Melott, A. Stebbins for useful discussions and F.R. Bouchet & L. Hernquist for letting us use their treecode. S.C. and S.D. thank the Aspen Center of Physics for its hospitality while some of this work was done; and S.D. thanks the Institute of Theoretical Physics in Santa Barbara where this work was completed. This research was supported in part by the National Science Foundation under Grant No. PHY94-07194, by the DOE, by NASA grant number NAG-5-2788 and by the Natural Sciences and Engineering Research Council of Canada.

## Appendix

It is often useful to have an analytic fit for the linear transfer functions calculated in Section 2. Since our models range from HDM to CDM and some care must be taken if a single functional form is to be used for all models. We choose analytic functions of the form

$$2 \log_{10} T(k) = \sum_{i=1}^{6} p_i \left( h^{-2} k \right)^{n_i}$$

where $k$ is measured in units of Mpc$^{-1}$ and $n_i = i/6$. The fitting functions are valid for $k \lesssim 0.5$ Mpc$^{-1}$ The values of the parameters $p_i$ for the models considered are given in the table 2.

Table 2:

| mass | $p_1$ | $p_2$ | $p_3$ | $p_4$ | $p_5$ | $p_6$ |
|---|---|---|---|---|---|---|
| $m_0$ | -13.73 | 112.0 | -345.9 | 505.6 | -348.7 | 85.18 |
| $2m_0$ | 0.4449 | -10.22 | 56.25 | -122.8 | 115.0 | -42.20 |
| $4m_0$ | -12.78 | 94.30 | -257.4 | 328.1 | -200.4 | 45.42 |
| $8m_0$ | 5.271 | -49.26 | 173.0 | -280.7 | 206.9 | -57.63 |

# References

Adams, F. C., Bond, J. R., Freese, K., Frieman, J. A., & Olinto, A. 1993, Phys. Rev., D47, 426
Albrecht, A. & Stebbins, A. 1992, Phys. Rev. Lett., 69, 2615
Babu, K. S., Rothstein, I. Z., & Seckel, D. 1993, Nuc. Phys., B403, 725
Barbieri, R. & Dolgov, A. 1990, Phys. Lett., B237, 440
Barbieri, R. & Dolgov, A. 1991, Nucl. Phys., B349, 742
Bardeen, J. M., Bond, J. R., Kaiser, N., & Szalay, A. S. 1986, ApJ, 304, 15
Barger, V., Deshpande, N., Pal, P. B., Phillips, R. J. N., Whisnant, K. 1991, Phys. Rev., D43, R1759
Bond, J. R. & Efstathious, G. 1991, Phys. Lett., B265, 245
Bond, J. R., Efstathiou, G., & Silk, J. 1980, Phys. Rev. Lett., 45, 1980
Bond, J. R. & Szalay, A. S. 1983, ApJ, 274, 443
Bond, J. R., Szalay, A. S. & Turner, M. S. 1982, Phys. Rev. Lett., 48, 1636
Bouchet, F. R., & Hernquist, L. 1988, ApJS, 68, 521
Braun, E., Dekel, A., & Shapiro, P. R. 1988, ApJ, 328, 34
Bunn, E. F., Scott, D., & White, M. 1994, astro-ph/9409003 (BSW)
Butler, M. N. & Malaney, R. A. 1992, Phys. Lett., B283, 298
Carlberg, R. G. 1988, ApJ 332, 26
Carlberg, R. G. 1991, ApJ 367, 385
Carlberg, R. G. 1994, ApJ 433, 468
Carlberg, R. G., & Dubinski, J. 1991, ApJ, 369, 13
Cen, R., & Ostriker, J. P. 1992, ApJ, 399, L113
Cline, J. M. 1992, Phys. Rev. Lett. 68, 3137
da Costa, L. N., Pellegrini, P. S., Davis, M., Meiksin, A., Sargent, W., & Tonry, J. L. 1991, ApJS, 75, 935
Copi, C. J., Schramm, D. N., & Turner, M. S., 1995, Science, 267, 192
Couchman, H. M. P., & Carlberg, R. G., 1992, ApJ, 389, 453
Cowsik, R. & McClelland, J. 1972, Phys. Rev. Lett., 29, 669
van Dalen, A. & Schaefer, R. K. 1992, ApJ, 398, 33
Davis, M., & Peebles, P. J. E., 1983, ApJ, 267, 465
Davis, M., Efstathiou, G., Frenk, C. S., & White, S. D. M. 1985, ApJ, 292, 371
Davis, M., Summers, F., & Schlegel, D. 1992, Nature, 359, 393
Dodelson, S., Gyuk, G., & Turner, M. S. 1994, Phys. Rev. Lett., 72, 3754
Dodelson, S. & Widrow, L. M. 1994, Phys. Rev. Lett., 72, 17
Dolgov, A. 1981, Sov. J. Nucl. Phys., 33, 700

Efstathiou, G., Bond, J. R., and White S. D. M. 1992, MNRAS, 258, 1p
Efstathiou, G., Frenk, C. S., White, S. D. M., & Davis, M. 1988, MNRAS, 235, 715 (EFWD)
Efstathiou, G. *et al.* 1990, Nature, 348, 705
Einasto, J. E., Klypin, A. A., & Saar, E. 1986, MNRAS, 219, 457
Enqvist, K., Kainulainen, K., & Maalampi, J. 1990a, Phys. Lett., B244, 186
Enqvist, K., Kainulainen, K., & Maalampi, J. 1990b, Phys. Lett., B249, 531
Enqvist, K., Kainulainen, K. & Thomson, M. 1992, Nucl. Phys., B373, 498
Frenk, C. S., White, S. D. M., Davis, M., Efstathiou, G. 1988, ApJ, 327, 507
Fry, J. N., & Gaztañaga, E. 1993, ApJ, 413, 447
Gelb, J. M., & Bertschinger, E. 1994, ApJ, 436, 491 (GB)
Gelb, J. M., Gradwohl, B.-A., & Frieman, J. A. 1993, ApJ, 403, L5
Geller, M. J., & Huchra, J. P. 1983, ApJS, 52, 61
Geller, M. J., & Huchra, J. P. 1989, Science, 246, 897
Gerstein, G. & Zel'dovich, Ya. B. 1966, Zh. Eksp. Teor. Fiz. Pis'ma Red., 4, 174
Gorski, K. M. *et al.* 1994, ApJ, 430, L89
Gott, J. R., Turner, E. L. 1977, ApJ, 216, 357
Hernquist, L., Bouchet, F.R., & Suto, Y. 1991, ApJS, 75, 231
Hockney, R. W., & Eastwood, J. W., 1981, Computer Simulation Using Particles (New York: McGraw Hill)
Holtzman, J.A. 1989, ApJS, 71, 1
Huchra, J. P., & Geller, M., 1982, ApJ, 257, 423
Huchra, J. P., Davis, M. J., Latham, D., & Tonry, J. 1983, ApJS, 52, 89
Huchra, J. P., Geller, M. J., de Lapparent, V., & Corwin, H. G. 1990, ApJS, 72, 433
Katz, N., Hernquist, L., & Weinberg, D. H. 1992, ApJ, 399, L109
Klypin, A., Holtzman, J., Primack, J., & Regös, E. 1993, ApJ, 416, 1
Kolb, E. W. & Turner, M. S. 1990, The Early Universe (Redwood City, CA: Addison-Wesley)
Langacker, P. 1989 University of Pennsylvania Report No. UPR 0401T (unpublished)
de Lapparent, V., Geller, M. J., & Huchra, J. P. 1986, ApJ, 302, L1
de Lapparent, V., Geller, M. J., & Huchra, J. P. 1991, ApJ, 343, 1
Malaney, R., Starkman, G., & Widrow, L. M. 1995 (in preparation)
Manohar, A. 1987, Phys. Lett., B186, 370
Marx, G. & Szalay, A. 1972, Neutrino '72 (eds. A. Frenkel & G. Marx, OMKDT-Technoinform, Budapest)
Mather, J. C. et al. 1994, ApJ 420, 439
Melott, A. L. 1987, MNRAS 228, 1001
Mikheyev, S. P. & Smirnov, A. Yu. 1986, Nuovo Cimento C9, 19
Mo, H. J., Jing, Y. P., & Börner, G. 1993, MNRAS, 264, 825
Moore, B., Frenk, C. S., & White, S. D. M. 1993, MNRAS, 261, 827 (MFW)
Moutarde, F., Alimi, J.-M., Bouchet, F. R., Pellat, R., & Ramani, A. 1991, ApJ, 382, 377
Nolthenius, R., Klypin, A., & Primack, J. 1994, preprint
Nolthenius, R., & White, S. D. M. 1987, MNRAS, 225, 505
Olive, K. A. & Turner, M. S. 1982, Phys. Rev., D25, 213
Pagels, H. R. & Primack, J. R. 1982, Phys. Rev. Lett., 48, 223
Peacock, J. A., & Dodds, S. J. 1994, MNRAS, 267, 1020 (PD)
Peebles, P. J. E. 1980, The Large Scale Structure of the Universe (Princeton: Princeton University Press)

Peebles, P. J. E. 1982, ApJ, 258, 415
Peebles, P. J. E. 1992, Principles of Physical Cosmology (Princeton: Princeton University Press)
Peebles, P. J. E. 1984, ApJ, 284, 439
Schechter, P., 1976, ApJ 203, 297
Shafi, Q & Stecker, F. W. 1984, Phys. Rev. Lett. 53, 1292
Smoot, G. F., et al. 1992, ApJ 396, L1
Strauss, M. A., Davis M., Yahil, A., Fisher, K. B., & Tonry, J. 1992, ApJS, 83, 29
Taylor, A. N., Rowan-Robinson, M. 1992, Nature, 359, 396.
Turner, M. S., Steigman, G., & Krauss, L. 1984 Phys. Rev. Lett., 52, 2090
Turner, M. S. 1991, Phys. Scr., T36, 167
Valdarnini, R. & Bonometto, S. A. 1985, A& A, 146, 235
Weinberg, D. H. 1994, Wide-Field Spectroscopy and the Distant Universe (eds. S. J. Maddox & A. Aragón-Salamanca, World Scientific, Singapore), in press
Weinberg, D. H., & Cole, S. 1992, MNRAS, 259, 652
Weinberg, S. 1972, Gravitation and Cosmology (New York: J. Wiley)
West, M. J., Villumsen, J. V., & Dekel, A. 1991, ApJ, 369, 287
White, S. D. M., Frenk, C. S., & Davis, M. 1983, ApJ, 274, L1
White, S. D. M., Frenk, C. S., Davis, M., & Efstathiou, G. 1987, ApJ, 313, 505
White, S. D. M., Davis, M., Efstathiou, G., & Frenk, C. S. 1987, Nature, 330, 451
Wolfenstein, L. 1978, Phys. Rev., D17, 2369
Zel'dovich, Ya. B. 1970, A&A, 5, 84
Zeng, N., & White, S. D. M. 1990, ApJ, 374, 1
Zurek, W. H., Quinn, P. J., Salmon, J. K., Warren, M. S. 1994, ApJ, 431, 559